\documentclass[useAMS,usenatbib]{mn2e}
\usepackage{captcont}
\usepackage{longtable}
\usepackage{graphicx}
\usepackage{natbib}
\usepackage{lscape}
\usepackage{rotating}
\usepackage{hyperref}
\usepackage{amssymb}
\usepackage{color}
\title[Variation of X-ray spectral index with {X-ray} Eddington ratio ]{Relationship between X-ray spectral index and {X-ray} Eddington ratio for Mrk~335 and Ark~564}
\author[R. Sarma et al.]{R. Sarma$^{1}$\thanks{e-mail: sharma.rathin@gmail.com}, S. Tripathi$^{2}$, R. Misra$^{2}$, G. Dewangan$^{2}$, A. Pathak$^{3}$ and J. K. Sarma$^{3}$\\
$^{1}$\textit{Department of Physics, Hojai College, Hojai, 782435, India}\\
$^{2}$\textit{Inter-University Centre For Astronomy and Astrophysics, Post Bag 4, Ganeshkind, Pune-411007, India }\\
$^{3}$\textit{Department of Physics, Tezpur University, 784028, India} }
\begin{document}
\date{}
\pagerange{\pageref{firstpage}--\pageref{lastpage}}
\maketitle
\label{firstpage}
\begin{abstract}
We present a comprehensive flux resolved spectral analysis of the
bright Narrow line Seyfert I AGNs, Mrk~335 and Ark~564  using observations by
XMM-Newton satellite. The mean and the flux resolved spectra are fitted by 
an empirical model consisting of two Comptonization components, one for
the low energy soft excess and the other for the high energy
power-law. A broad Iron line and a couple of low energies edges are
required to explain the spectra. For Mrk~335, the 0.3 - 10 keV luminosity relative to the  Eddington value, L{$_{X}$}/L$_{Edd}$, varied from 0.002 to 0.06. The index variation can be empirically described as $\Gamma$ = 0.6 log$_{10}$ L{$_{X}$}/L$_{Edd}$ + 3.0 for $0.005 <  L{_{X}}/L_{Edd} < 0.04$. At $ L_{{X}}/L_{Edd} \sim 0.04$ the spectral index changes and then continues to follow $\Gamma$
= 0.6 log$_{10}$ L$_{{X}}$/L$_{Edd}$ + 2.7, i.e. on a parallel track. We confirm
that the result is independent of the specific spectral model used by
fitting the data in the 3 - 10 keV band by only a power-law and an Iron
line. For Ark~564, the index variation can be empirically described as $\Gamma$ =
0.2 log$_{10}$ L$_{{X}}$/L$_{Edd}$ + 2.7 with a significantly large scatter as compared
to Mrk~335. Our results indicate that for Mrk~335, there may be accretion disk geometry changes which lead
to different parallel tracks.   {These changes could be related to structural changes in the
corona or enhanced reflection at high flux levels.} There does not seem to be any homogeneous or universal relationship
for the X-ray index and luminosity for different AGNs or even for the same AGN. 
\end{abstract}
\section{Introduction}

Active Galactic Nuclei (AGN) are known to emit  X-rays
as a result of physical processes active in the innermost regions near the central super-massive black hole. X-ray spectra of a typical AGN in the $2$-$10$ keV range show primarily the signature of a power-law continuum and an iron line. At energies $< 2 $~keV there is often a
soft excess over the power-law emission. The power-law emission 
is widely considered as the outcome of inverse
Compton scattering of thermally produced accretion disk seed
optical/UV photons by a corona of hot electrons close to the disc
\citep{Haardt1991, Haardt1993, Zdz2000}. However, the geometry and
size of the corona as well as the physical mechanism governing the energy
transfer between the two phases are not well understood.

While, the average value of photon index of the primary power-law for
AGN has been found to be $\Gamma \sim 1.9$, there is a large variation with $\Gamma$ ranging from $1.5$-$2.5$ \citep{Nandra1994, Page2005}. For
a given AGN, $\Gamma$ varies as a function of time. Understanding the 
behaviour of $\Gamma$ with other source properties is expected to give
important insight into the nature of the radiative mechanism and the
geometry of the inner regions.

Previous studies have shown the existence of a correlation
between the X-ray photon index $\Gamma$ and the source flux e.g.,
\citep{Perola et al.86, Singh1991, Mushotzky et al.93, Done2000,
Nandra2001}. For many Seyfert 1 AGNs, the power-law index shows
significant variation and generally follows the trend of steeper
$\Gamma$ with increasing source intensity. 
 During 1997, MCG~6-30-15 was observed with \textit{RXTE} for a
duration of 8 days $\sim$ 910~ks. {The range of $\Gamma$ observed was 1.8-2.2 with flux variation of $\sim$~3.3 in the energy band 2-10 keV.} This observation showed a tight correlation between the photon index $\Gamma$ and the
source flux in the sense that the power-law steepens as the source
gets brighter \citep{Vaughan2001}. The steepening of $\Gamma$ {(range 1.89-2.34)} with
flux has also been observed in IRAS 13224-3809, where a 10 day long
\textit{ASCA} observation revealed a change in the
power-law flux by a factor of $\sim$~3.2 in the 2-10 keV range
\citep{Dewangan et al.2002}. \textit{RXTE} observed NGC~5548 on five
occasions in 1998 during which a clear positive correlation
between the photon index {(range 1.75-1.93)}and the 2-10 keV flux was seen {with flux variation of $\sim$~1.84.}\citep{Chiang
et al.2000}. {For the soft energy band (0.3-2.0 keV), similar trend between photon index (range 1.7-2.0) and flux was observed for a 2 day long \textit{BEPPOSAX} observation of 3C 120 with flux variation of  $\sim$~1.5 \citep{Zdziarski2001}.} The Seyfert galaxy NGC 4151 was observed on seven occasions
by \textit{Ginga} during May 1987-January 1989 ($\sim$21 months). In
this observation significant flux variation {($\sim$~3)} has been seen along with
the correlation between the photon index and flux \citep{Yaqoob1991}.
The 2009 \textit{XMM-Newton} observation of Mrk~335 has also shown
`softer when brighter' property \citep{Grupe et al.2012}.
{In all the above examples of AGNs, the flux variation during the observations does not exceed a factor of $\sim 4$. }

Studies of a sample of AGNs have also provided evidence that there is significant positive correlation between the X-ray photon
index and the {bolometric} Eddington ratio, $\lambda = L_{Bol}/L_{Edd}$, where
L$_{Edd}$ is the Eddington luminosity \citep{Laor et al.1997, Lu1999, Wang2004, Shemmer et
al.2006}. With larger samples involving sources with higher redshifts and greater
luminosities, it was observed that the hard X-ray photon index
correlates with the {bolometric} Eddington ratio when $\lambda \gtrsim 0.01$
\citep{Porquet et al.2004, Shemmer et al.2006, Saez et al. 2008,
Sobolewska2009, Cao 2009, Zhou2010, Liu et al. 2012, Brightman2013}
but when  $\lambda <$ 0.01, especially for low-luminosity
AGNs, anti-correlation is seen between $\Gamma$ and $\lambda$ \citep{Gu2009, 
Constantin et al. 2009, Younes et al. 2011, Xu 2011}. In general, there
have been several studies to investigate the relation between the photon index and
$\lambda$ for different samples of AGN \citep[e.g.][]{Risaliti2009, Jin2012}.
Studies incorporating \textit{ROSAT} and \textit{ASCA}
observations showed the correlation between $\Gamma$ and full width at
half maximum (FWHM) of the H$\beta$ emission line \citep{ Boller1996,
Brandt1997, Dewangan et al.2002}. 

There is a need to study the relation between spectral index and
 Eddington ratio spanning a larger range in $L/L_{Edd}$. This
is provided by \textit{XMM-Newton} data for the 
bright and highly variable Narrow Line Seyfert 1 (NLS1) galaxy
Mrk~335. For comparison we also analyse the extensive data available for 
another NLS1, Ark~564.

{Narrow line Seyfert 1 galaxies (NLS1) form a subset
of AGN which exhibit exceptional features in terms of emission-line
and continuum properties. Unlike Seyfert 2 galaxies which show narrow
optical emission lines, NLS1 show the broad emission-line optical
spectra of Seyfert 1 galaxies, but with the narrowest Balmer lines
from the broad line region (full width at half maximum (FWHM)
$\lesssim$ 2000~ km~s$^{-1}$ with relatively weak [OIII]$\lambda$5007
emission) and prominent optical Fe II emission \citep{Ost1985,
Veron2001}. NLS1 also show a number of other extreme properties in
X-rays e.g., strong soft excess emission below 1 keV, steep 2-10 keV
power-law continuum and very rapid and large X-ray variability
\citep{Leighly1999, Gallo2004}. Their X-ray spectra often present
complex behaviour with the presence of cold and ionised absorption,
partial covering and reflection components (\citep{Komossa2008} and
references therein). Also NLS1 galaxies as a class follow the
$M_{BH}-\sigma_{\ast}$ relation if the widths of emission lines not
strongly affected by outflow components are used as a surrogate for
$\sigma_{\ast}$ \citep{Komossa2007}. Due to all these extreme and
ambiguous properties, NLS1 as a special class of AGN seem to challenge
the Unified model and need a more careful investigation. Recent
observations and surveys have pointed towards the presence of smaller
black hole masses in NLS1 galaxies \citep{Barth2005, Botte et al.2004} and
which could possibly represent an important connection with the less
explored intermediate mass black holes. It is believed that the
extreme X-ray and other observed properties of NLS1 galaxies may be due to
an extreme value of a fundamental physical parameter related to the
accretion process. This fundamental parameter is most likely to  be the
accretion rate relative to the Eddington rate. This is well supported by
more recent studies, theoretical considerations and by black hole mass
estimates from optical emission-line and continuum measurements that
NLS1 are accreting close to their Eddington rates \citep{Boller1996,
Boroson2002, Xu2003, Grupe2004, Warner2004, Collin2006} and hence
should be considered important testbeds of accretion models. Recent
X-ray study of NLS1 using an optically selected SDSS sample have shown
that some NLS1 show steep X-ray spectra and strong Fe II emission
while some do not. In this study, a strong correlation was also found
between $\Gamma$ and the luminosity at 1 keV, $L_{1keV}$ suggesting differences in
$L_{bol}/L_{Edd}$ among the NLS1s in the sample \citep{Williams2002,
Williams2004}. }

Mrk~335, also known as PG003+199 is a nearby NLS1 galaxy at a
redshift z=0.026 \citep{Longinotti et al.07a} and has a well measured
black hole mass of 1.4 $\times$10$^{7}$~M$_{\odot}$ from reverberation
mapping \citep{Peterson et al.2004, Grier et al.2012}. It 
has been the target of most X-ray observatories. \textit{XMM-Newton}
observed Mrk~335 for the first time in 2000. Analysing the RGS data, an
absorption edge at 0.54 keV due to Galactic oxygen was reported and
the soft excess was described as a combination of bremsstrahlung
emission and ionised reflection from the accretion disk \citep{Gondoin
et al.02}. The \textit{XMM-Newton} observation was later reanalysed
and a narrow absorption feature at 5.9 keV was found \citep{Longinotti
et al.07a}. In 2006 January, \textit{XMM-Newton} re-observed Mrk~335
for 133~ks which revealed a double-peaked Fe emission
feature with peaks at 6.4 and 7.0 keV \citep{O'Neill et al. 2007}.
\citep{Lar08} studied Mrk~335 using a 151~ks \textit{Suzaku}
observation performed in 2006. They modelled the data using a power law and two reflectors in
which an ionised, heavily blurred, inner reflector produces most of
the soft excess, while an almost neutral outer reflector (outside
$\sim$ 40r$_{g}$) produces most of the Fe line emission. They
also verified their model using the 2006 \textit{XMM-Newton} data and
did not see any correlation between photon index and power law
flux. But subsequently a marginal trend, i.e., the source becomes
softer with increasing count rate  has been
reported for the same data \citep{Grupe et al.2012}. When observed by \textit{XMM-Newton} in July 2007 for 22~ks, Mrk~335 was
found to be in extremely low X-ray flux state \citep{Grupe et
al.08a}. The spectrum of this low flux state of Mrk~335 was explained
by partial covering and blurred reflection
models.   Mrk~335 was
again observed by \textit{XMM-Newton} in 2009 for 200~ks spread over
two consecutive orbits. The X-ray continuum and timing properties have
been described by using a blurred reflection model \citep{Gallo et
al.13}. The 2009 \textit{XMM-Newton} data have also been analysed by
\citet{Grupe et al.2012} along with the \textit{Swift} data. Partial
absorption and blurred reflection models gives equally good fit to the
spectrum. The number of observations of Mrk~335 by \textit{XMM-Newton}
over time provides an opportunity to study the variation of the high energy photon index with luminosity in a systematic manner.

 At a redshift z=0.02469 \citep{Huchra et al.1999}, Ark~564 is the
brightest NLS1 galaxy in the 2.0-10.0 keV range, L$_{(2-10)keV}$ =
2.4 $\times$ 10$^{43}$ erg s$^{-1}$ \citep{Turner et
al.2001}. Ark~564 has been studied across all wavebands
\citep{Shemmer et al.2001, Romano et al.2004}. In the 2000 and 2001
\textit{XMM-Newton} observations of Ark~564, \citet{Vig2004} reported an 
edge-like feature in the EPIC data at $\sim$ 0.73~keV and interpreted it as the 
OVII K absorption edge. The $\sim$100~ks 2005 \textit{XMM-Newton} observation of Ark~564 have been described as
either a power law and two black-bodies or a relativistically blurred
photo ionised disk reflection model \citep{Papadakis et
al.2007}. \citet{Dewangan et al. 2007} analysed the 2005 data and
found two warm absorber phases. During 2011, Ark~564 has been observed
by \textit{XMM-Newton} for eight occasions. \citet{Legg et
al.2012} confirmed a significant soft lag in the $0.3$-$1.0$ keV and
$4.0$-$7.5$ keV bands and suggested a distant reflection origin. Thus,
Ark~564 has been also observed several times by \textit{XMM-Newton} making
it another good candidate.

 Here we investigate the dependence of power-law index with the
 {X-ray} Eddington ratio for the combined EPIC spectra of Mrk~335 and
 Ark~564. Using flux resolved spectroscopy we study the power law
 index variation against {X-ray} Eddington ratio. The paper is organised as
 follows. Section 2 describes the observations and the data reduction
 procedure and  Section 3  explains the spectral model used for
 the analysis. The details of the flux resolved spectroscopy
 are given in section 4. Finally, we conclude this work with a summary of the results
 and discussion in section 5.

\section{Observations and data reduction}

 \begin{table}
\label{tabobs}
\begin{center}

\caption{Observation log of Mrk~335 and Ark564 by \textit{XMM-Newton}}
\begin{tabular}{ccccc}
\hline
\multicolumn{4}{c}{Mrk~335} \\
\hline
OBS. ID& Duration(s)&Date& Pile-up \\
\hline
0101040101&36910&2000-12-25 &yes\\
0306870101&133251&2006-01-03&no \\
0510010701&22580&2007-07-10 &no\\
0600540601&132315&2009-06-11&no \\
0600540501&82615&2009-06-13 &no\\
\hline
\multicolumn{4}{c}{Ark~564} \\
\hline
0006810101&34466&2000-06-17&no\\
0006810301&16211&2001-06-09&no\\
0206400101&101774&2005-01-05&no\\
0670130201 &59500 &2011-05-24&yes \\
0670130301 &55900 &2011-05-30&no \\
0670130401 &63620 &2011-06-05 &no\\
0670130501 &67300 &2011-06-11 &yes\\
0670130601 &60900 &2011-06-17&no \\
0670130701 &64420 &2011-06-25&no\\
0670130801 &58200 &2011-06-29&yes \\
0670130901 &55900 &2011-07-0&yes \\
\hline 
\end{tabular}
\end{center}

\end{table}

For the analysis, we use all the available archival data of Mrk~335 and
Ark~564 from the \textit{XMM-Newton} observatory \citep{Jansen et al.2001}. The
list of observations are shown in Table 1. The \textit{XMM-Newton} data have been processed
in the standard way using the SAS version 12.0. For all cases we have
considered data from the EPIC-pn camera \citep{Struder et al.01} only.

 The data were cleaned for high background flares and were selected
 using the conditions PATTERN $\leqq{4}$ and FLAG == 0. We have
 checked for pile-up in all cases using SAS task \textit{epatplot} and
 found that for both the sources some observations are affected by
 pile-up (Table 1). We reduced the pile-up  by excluding the
 innermost source emission using an annular region as considered by \citep{Legg et
 al.2012}. 
 Except for the pile-up affected observations, the source spectra have been extracted from
 circular regions of radius 35 arc s centred on the maximum source
 emission. The redistribution matrices and auxiliary response files were created
 by the SAS task \textit{especget}. Spectra were grouped such that each
 bin contained at least 30 counts. Spectral fits to the data were
 performed with XSPEC version 12.8.0 \citep{Arnaud96}. The background
 subtracted light curves have been produced with the tool \textit{EPICLCCORR}
 which are binned with 400~s bins.

\begin{figure*}
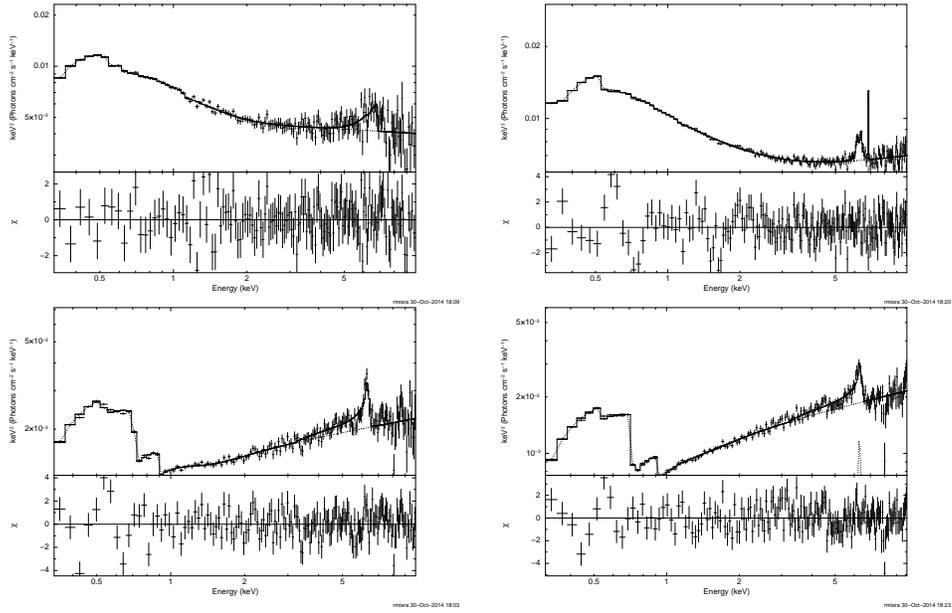

\begin{centering}
\includegraphics[width = 4cm, angle=270]{spec0101040101.ps}
\includegraphics[width = 4cm, angle=270]{spec0306870101.ps}
\includegraphics[width = 4cm, angle=270]{spec0600540501.ps}
\includegraphics[width = 4cm, angle=270]{spec0600540601.ps}
\caption{{EPIC PN unfolded spectra and residuals ($\Delta \chi$) of Mrk~335 observations i.e. 0101040101 
(top right), 0306870101 (top left), 0600540501 (bottom right) and 0600540601 (bottom left) for
the phenomenological model with best fit parameters listed in Table \ref{mrk335spec}}. }

\label{sm}
\end{centering}
\end{figure*}

\begin{figure*}
\includegraphics{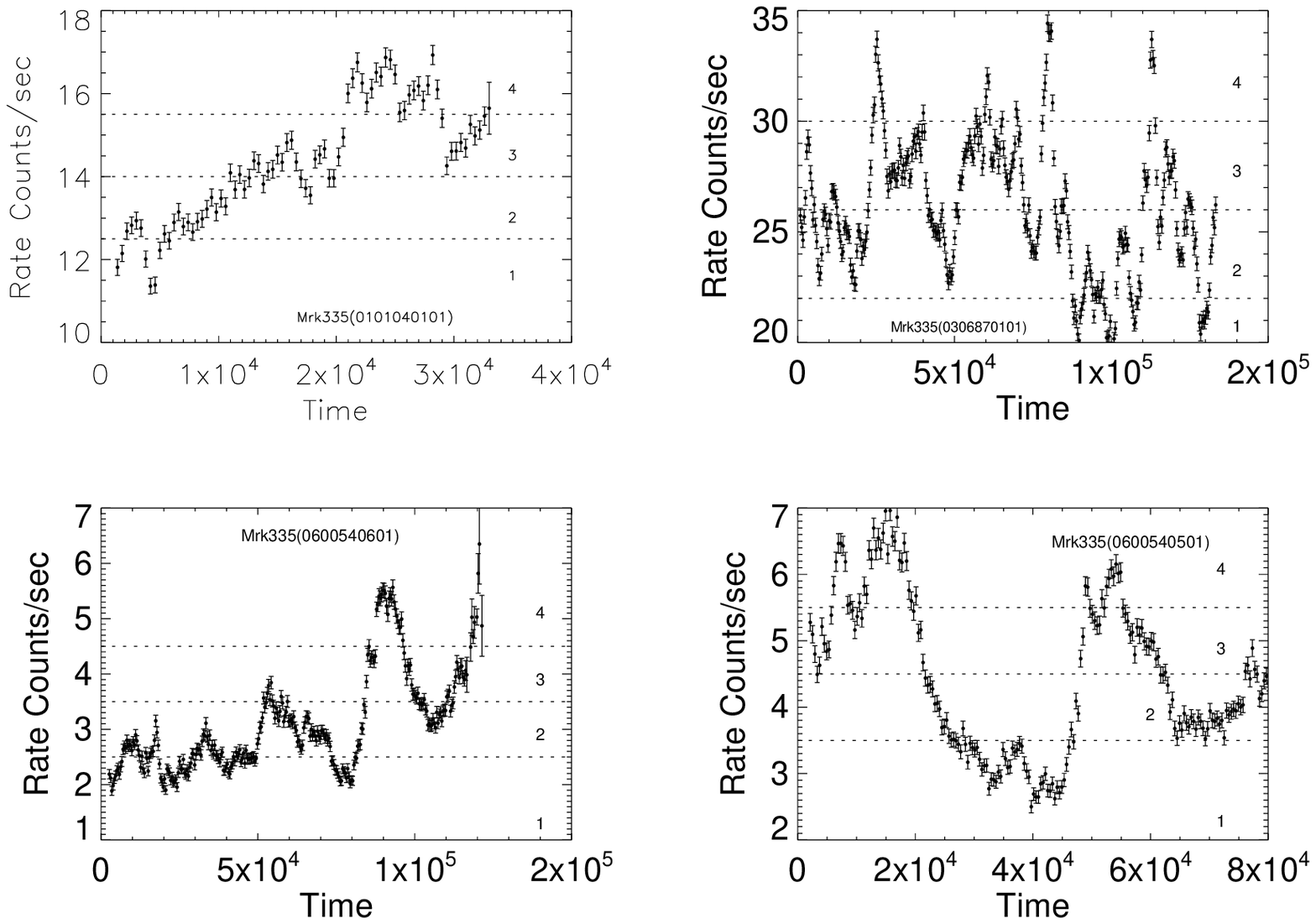}
\caption{EPIC PN light curve (in 400~s bins) of Mrk~335.
 The light curves were extracted in the energy range of (0.3-10.0) keV. 
The time ranges for flux resolved spectroscopy have been selected corresponding 
to different counts s $^{-1}$ as shown by the horizontal dotted lines.}
\label{lc1}
\end{figure*}

\begin{figure*}
\includegraphics{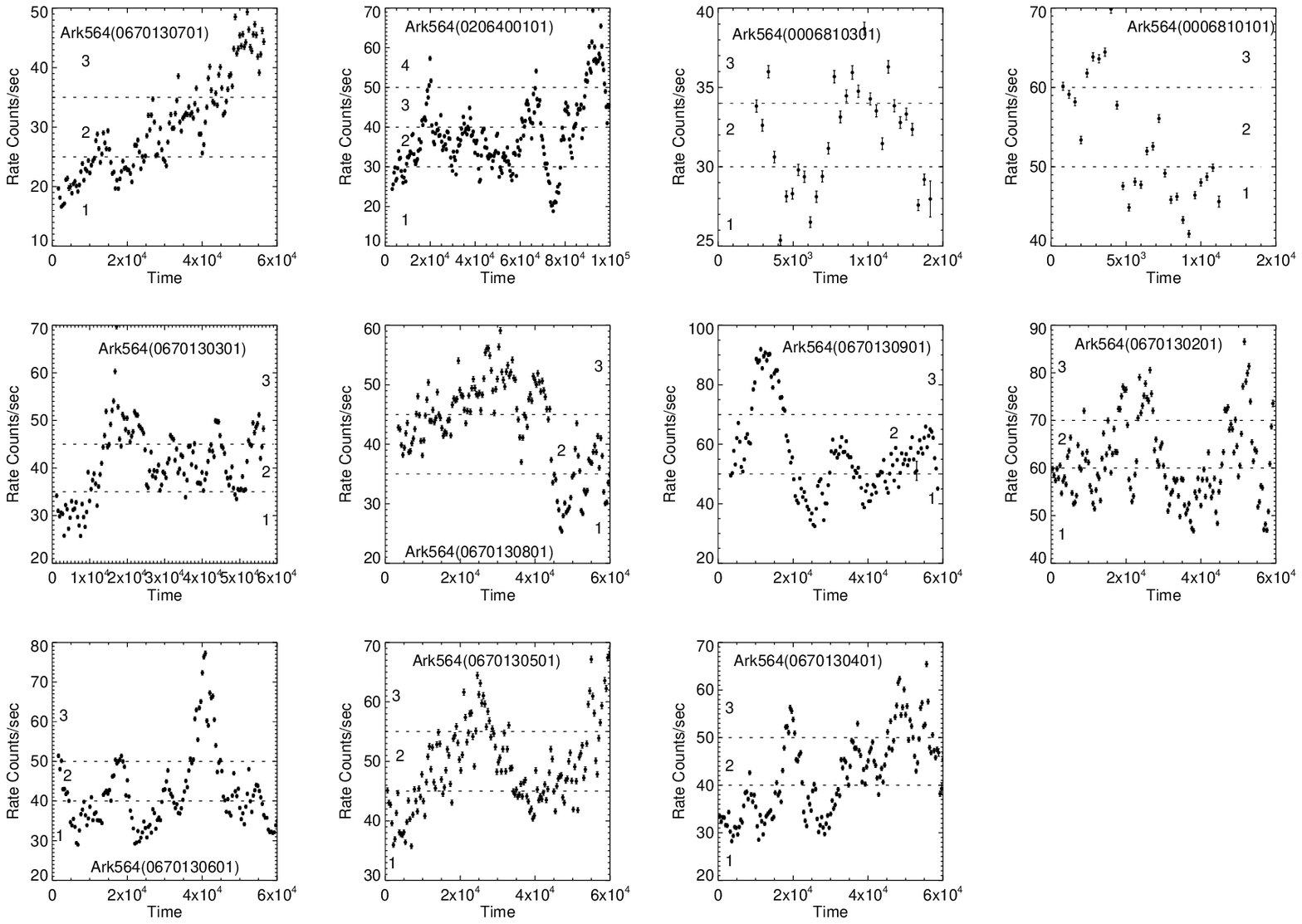}
\caption{EPIC PN light curve (in 400~s bins) of Ark~564.
 The light curves were extracted in the energy range of (0.3-10.0) keV. 
The time ranges for flux resolved spectroscopy have been selected corresponding 
to different counts s $^{-1}$ as shown by the horizontal dotted lines.}
\label{lc2}
\end{figure*}

\section{PHENOMENOLOGICAL MODEL}

We fit the EPIC-pn spectral data in the $0.3$-$10.0$ keV range, by a simple phenomenological model consisting
of two thermal Comptonization models. In particular, the soft excess is described by the
XSPEC model ``nthComp" and for the hard X-ray emission, we have used
XSPEC convolution model ``Simpl" \citep{Steiner et al.08}. 
 Galactic
absorption is taken to be  N$_{H}$=3.99$\times$10$^{20}$
cm$^{-2}$ for Mrk~335 and 
N$_{H}$=5.34$\times$10$^{20}$ cm$^{-2}$ for Ark~564 \citep{Kalberla et al.05}.
For Mrk~335 there are clear residuals at around $\sim 6$ keV signifying the
presence of a broad Iron line which we modelled using the ``diskline'' \citep{Fabian et al. 1989}. The inner radius of the line emitting region R$_{in}$ is
fixed at 6r$_{g}$ and the outer radius r$_{out}$ is fixed at
1000r$_{g}$. The inclination of the disk is fixed at 40$^{\circ}$. For the
``nthcomp'' model we find that the results are not sensitive to the
input seed photons which we fix to be a blackbody at $0.05$ keV. For both the
sources, addition of 2 or 3 low energy absorption edges improves the spectral fitting. For
one of the Mrk~335 observations there is also a hint of a narrow emission line at $7$ keV. {Figure \ref{sm} presents the unfolded spectra for the observations 0101040101, 0306870101, 0600540501 and 0600540601 of Mrk~335.}
The best fit spectral parameters and the reduced $\chi^2$ for all the observations are listed
in Tables \ref{mrk335spec} and \ref{ark564spec}. 

\begin{table*}
\begin{center}
\caption{Spectral parameters for Mrk~335 derived from different \textit{XMM-Newton} observations in the (0.3-10.0)keV range.} 
\label{mrk335spec}
\begin{tabular}{lcccccc}
\hline
\hline
Model & parameter&0101040101&0306870101&0510010701&0600540601 &0600540501 \\ 
zedge1& $E_c$ (keV)& 0.63(froze)&$0.361_{-0.004 }^{+0.003}$ &$  1.047_{-  0.035 }^{+  0.032}$ &$  0.730_{-  0.003 }^{+  0.004}$ &$  0.730_{-  0.004 }^{+  0.004}$\\
&$\tau$& $  0.097_{-  0.031 }^{+  0.028}$&$0.228_{-0.006 }^{+0.027}$ &$  0.369_{-  0.093 }^{+  0.091}$ &$  0.776_{-  0.042 }^{+  0.027}$ &$  0.553_{-  0.017 }^{+  0.029}$\\
zedge2&$E_c$ (keV)& $  1.127_{-  0.046 }^{+  0.042}$ &$1.120_{-0.230 }^{+0.034}$ &$  1.509_{-  0.063 }^{+  0.060}$ &$  0.943_{-  0.012 }^{+  0.011}$ &$  0.922_{-  0.011 }^{+  0.014}$\\
&$\tau$& $  0.063_{-  0.031 }^{+  0.031}$ &$0.026_{-0.006 }^{+0.006}$ &$  0.377_{-  0.095 }^{+  0.101}$ &$  0.313_{-  0.041 }^{+  0.030}$ &$  0.238_{-  0.040 }^{+  0.024}$\\
zedge3&$E_c$ (keV)&-&-&0.686(Froze) &-&-\\
&$\tau$&-& -& $  0.524_{-  0.100 }^{+  0.104}$  &-&-\\
Diskline$^\dagger$&$E_c$ (keV)& $  6.313_{-  0.184 }^{+  0.362}$ &$6.275_{-0.041 }^{+0.032}$&$  6.009_{-  0.068 }^{+  0.056}$ &$  6.022_{-  0.061 }^{+  0.050}$ &$  5.952_{-  0.064 }^{+  0.070}$\\
&$\beta$& $ -2.609_{-  5.900 }^{+  0.749}$ & $ -1.326_{-0.414 }^{+0.643}$ &$ -9.668_{-  8.006 }^{+  3.607}$ &$ -7.416_{-  4.753 }^{+  2.297}$ &$ -5.061_{-  4.475 }^{+  1.497}$ \\
&Norm($\times$10$^{-4}$)&$  0.449_{-  0.168 }^{+  0.254}$ & $  0.217_{-  0.038 }^{+  0.044}$ &$  0.549_{-  0.076 }^{+  0.074}$   &$  0.276_{-  0.033 }^{+  0.034}$ & $  0.292_{-  0.022 }^{+  0.042}$\\
zgaussian&$E_c$ &-&$7.004_{-0.031 }^{+0.030}$&-&-&-\\
&Norm($\times$10$^{-3}$)&-&$ 0.007_{-  0.002 }^{+  0.002}$&-&-&-\\
Simpl&$\Gamma$& $  2.094_{-  0.049 }^{+  0.024}$ & $  1.885_{-  0.005 }^{+  0.009}$ &$  1.064_{-  0.046 }^{+  0.073}$ &$  1.676_{-  0.007 }^{+  0.018}$ &$  1.787_{-  0.008 }^{+  0.018}$\\
&FracScat&$  0.136_{-  0.009 }^{+  0.011}$  & $  0.087_{-  0.001 }^{+  0.002}$  & $  0.310_{-  0.104 }^{+  0.324}$&$  0.233_{-  0.007 }^{+  0.012}$ &$  0.193_{-  0.005 }^{+  0.007}$\\
nthComp&$\Gamma$&$  2.862_{-  0.057 }^{+  0.052}$  & $  3.277_{-  0.023 }^{+  0.008}$  &$  1.980_{-  0.092 }^{+  0.094}$ & $  1.922_{-  0.213 }^{+  0.087}$  &$  2.043_{-  0.219 }^{+  0.069}$ \\
&kT(keV)&$  0.266_{-  0.031 }^{+  0.051}$ & $  0.823_{-  0.020 }^{+  0.019}$  &$  0.210_{-  0.019 }^{+  0.019}$ & $  0.153_{-  0.018 }^{+  0.008}$&$  0.159_{-  0.020 }^{+  0.008}$\\
&Norm($\times$10$^{-2}$) &$  0.428_{-  0.032 }^{+  0.029}$ &$  0.705_{-  0.037 }^{+  0.007}$ & $  0.093_{-  0.039 }^{+  0.084}$ &$  0.060_{-  0.008 }^{+  0.003}$ & $  0.091_{-  0.009 }^{+  0.006}$\\
reduced $\chi^{2}$/d.o.f.&&  1.04/144&1.48/163  &1.70/130&1.40/161&1.52/159\\
\hline
\end{tabular}
\end{center}
{$^\dagger$The values of r$_{in}$ and r$_{out}$ have been fixed at $6 r_g$ and $1000 r_g$ respectively while
 the inclination has been fixed at 40$^{\circ}$.}
\end{table*}

\begin{table*}
\label{ark564spec}
\begin{center}
\caption{Spectral parameters for Ark~564 
derived from different \textit{XMM-Newton} observations in the (0.3-10.0)keV range.}
\begin{tabular}{lccccccc}
\hline
\hline
 Model & Parameter&0006810101& 0670130301&0670130801&0670130901&0670130201&0670130601\\
zedge1& $E_c$ (keV)&$  0.508_{-  0.015 }^{+  0.013}$ &$  0.708_{-  0.008 }^{+  0.008}$&$  0.345_{-  0.015 }^{+  0.013}$ & $  0.723_{-  0.010 }^{+  0.010}$ &$  0.330_{-  0.018 }^{+  0.011}$ &$  0.347_{-  0.006 }^{+  0.005}$\\
&$\tau$&$  0.111_{-  0.022 }^{+  0.022}$ &$  0.105_{-  0.010 }^{+  0.012}$ &$  0.244_{-  0.051 }^{+  0.050}$ &$  0.093_{-  0.013 }^{+  0.014}$ &$  0.359_{-  0.044 }^{+  0.274}$ &$  0.269_{-  0.021 }^{+  0.044}$\\
zedge2&$E_c$ (keV)&$  0.712_{-  0.013 }^{+  0.011}$ &$  1.054_{-  0.045 }^{+  0.041}$ &$  0.713_{-  0.006 }^{+  0.012}$ &$  1.071_{-  0.062 }^{+  0.053}$ &$  0.751_{-  0.023 }^{+  0.020}$ &$  0.714_{-  0.012 }^{+  0.011}$\\
&$\tau$&$  0.147_{-  0.022 }^{+  0.030}$ &$  0.033_{-  0.014 }^{+  0.013}$ &$  0.072_{-  0.010 }^{+  0.013}$ &$  0.029_{-  0.011 }^{+  0.019}$ &$  0.045_{-  0.007 }^{+  0.013}$ &$  0.063_{-  0.012 }^{+  0.012}$\\
zedge3&$E_c$ (keV)&$  1.122_{-  0.094 }^{+  0.126}$ &$  1.231_{-  0.039 }^{+  0.038}$ &$  1.160_{-  0.018 }^{+  0.019}$ &$  1.231_{-  0.087 }^{+  0.159}$ &$  1.217_{-  0.031 }^{+  0.033}$ &$  1.127_{-  0.019 }^{+  0.021}$\\
&$\tau$&$  0.022_{-  0.022 }^{+  0.023}$ & $  0.047_{-  0.014 }^{+  0.015}$ &$  0.086_{-  0.007 }^{+  0.013}$ &$  0.023_{-  0.018 }^{+  0.020}$ &$  0.057_{-  0.014 }^{+  0.014}$ &$  0.072_{-  0.011 }^{+  0.011}$\\
Simpl&$\Gamma$&$  2.474_{-  0.023 }^{+  0.024}$ &$  2.513_{-  0.013 }^{+  0.014}$ &$  2.418_{-  0.008 }^{+  0.014}$ &$  2.495_{-  0.017 }^{+  0.018}$ &$  2.448_{-  0.009 }^{+  0.019}$ &$  2.428_{-  0.014 }^{+  0.015}$\\
&FracScat&$  0.193_{-  0.007 }^{+  0.008}$ &$  0.192_{-  0.005 }^{+  0.004}$ &$  0.153_{-  0.004 }^{+  0.004}$ &$  0.194_{-  0.006 }^{+  0.006}$ &$  0.152_{-  0.005 }^{+  0.005}$ &$  0.151_{-  0.004 }^{+  0.004}$\\
nthComp&$\Gamma$&$  2.295_{-  0.047 }^{+  0.052}$ &$  2.292_{-  0.019 }^{+  0.010}$ &$  2.676_{-  0.050 }^{+  0.059}$ &$  2.353_{-  0.022 }^{+  0.016}$ &$  2.795_{-  0.055 }^{+  0.063}$ &$  2.862_{-  0.044 }^{+  0.051}$\\
&kT(keV)&$  0.184_{-  0.020 }^{+  0.020}$&$  0.195_{-  0.008 }^{+  0.009}$ &$  0.220_{-  0.020 }^{+  0.027}$&$  0.192_{-  0.010 }^{+  0.011}$&$  0.240_{-  0.023 }^{+  0.032}$ &$  0.254_{-  0.022 }^{+  0.031}$ \\
&Norm&$  0.014_{-  0.0009 }^{+  0.0008}$& $  0.012_{-  0.0003 }^{+  0.0003}$ &$  0.011_{-  0.0003 }^{+  0.0004}$&$  0.013_{-  0.0003 }^{+  0.0004}$ &$  0.015_{-  0.0004 }^{+  0.0005}$&$  0.011_{-  0.0003 }^{+  0.0004}$ \\
 red $\chi^{2}$/d.o.f.&&1.03/149&1.58/161&1.62/160&1.17/158&1.03/159&1.30/164\\
\hline
 Model & Parameter&0670130501&0006810301&0206400101&0670130701&0670130401\\
zedge1& $E_c$ (keV)&$  0.346_{-  0.007 }^{+  0.006}$ &$  0.338_{-  0.007 }^{+  0.006}$ &$  0.513_{-  0.006 }^{+  0.006}$ &$  0.587_{-  0.037 }^{+  0.045}$ &$  0.341_{-  0.018 }^{+  0.008}$\\
&$\tau$&$  0.406_{-  0.044 }^{+  0.039}$&$  0.595_{-  0.226 }^{+  0.226}$ & $  0.093_{-  0.009 }^{+  0.008}$ &$  0.035_{-  0.020 }^{+  0.019}$ &$  0.303_{-  0.041 }^{+  0.041}$\\
zedge2&$E_c$ (keV)&$  0.718_{-  0.009 }^{+  0.012}$ &$  0.489_{-  0.012 }^{+  0.012}$ &$  0.706_{-  0.007 }^{+  0.005}$ &$  0.716_{-  0.011 }^{+  0.013}$ &$  0.710_{-  0.005 }^{+  0.009}$\\
&$\tau$&$  0.069_{-  0.011 }^{+  0.011}$ &$  0.188_{-  0.017 }^{+  0.024}$ &$  0.116_{-  0.012 }^{+  0.011}$ &$  0.126_{-  0.024 }^{+  0.017}$ &$  0.075_{-  0.012 }^{+  0.012}$\\
zedge3&$E_c$ (keV)& $  1.197_{-  0.020 }^{+  0.023}$ &$  0.710_{-  0.017 }^{+  0.015}$ &$  1.111_{-  0.019 }^{+  0.018}$ &$  1.143_{-  0.027 }^{+  0.023}$ &$  1.164_{-  0.032 }^{+  0.030}$\\
&$\tau$& $  0.068_{-  0.012 }^{+  0.012}$ &$  0.102_{-  0.016 }^{+  0.017}$ &$  0.054_{-  0.009 }^{+  0.009}$ &$  0.061_{-  0.014 }^{+  0.014}$ &$  0.046_{-  0.011 }^{+  0.011}$\\
Simpl&$\Gamma$&$  2.437_{-  0.014 }^{+  0.011}$ &$  2.376_{-  0.012 }^{+  0.009}$ &$  2.472_{-  0.004 }^{+  0.009}$ &$  2.436_{-  0.016 }^{+  0.016}$ &$  2.416_{-  0.015 }^{+  0.012}$\\
&FracScat&$  0.144_{-  0.004 }^{+  0.004}$ &$  0.085_{-  0.015 }^{+  0.015}$ &$  0.183_{-  0.003 }^{+  0.003}$ &$  0.183_{-  0.005 }^{+  0.004}$ &$  0.159_{-  0.004 }^{+  0.005}$\\
nthComp&$\Gamma$&$  2.892_{-  0.029 }^{+  0.029}$ &$  3.127_{-  0.021 }^{+  0.024}$ &$  2.381_{-  0.022 }^{+  0.019}$ &$  2.175_{-  0.027 }^{+  0.022}$ &$  2.748_{-  0.051 }^{+  0.049}$\\
&kT(keV)&$  0.250_{-  0.023 }^{+  0.032}$&$  0.240_{-  0.004 }^{+  0.006}$ &$  0.188_{-  0.008 }^{+  0.009}$ &$  0.191_{-  0.008 }^{+  0.009}$ &$  0.233_{-  0.022 }^{+  0.032}$\\
&Norm&$  0.012_{-  0.0002 }^{+  0.0004}$ & $  0.009_{-  0.0001 }^{+  0.00003}$ &$  0.010_{-  0.0003 }^{+  0.0001}$ &$  0.009_{-  0.0003 }^{+  0.0002}$ &$  0.011_{-  0.0003 }^{+  0.0003}$\\
 red $\chi^{2}$/d.o.f.&&1.35/163&1.16/148&1.62/165&1.51/163&1.49/164\\
\hline
\end{tabular} 
\end{center}
\end{table*}

The reduced $\chi^2$ of the fits to the observations range from $1$ to $1.6$, which suggests
that the underlying spectra may be more complex than the phenomenological model used here, especially
at  low energies.  However for the motivation
of this work, the phenomenological model used here is adequate since even for a reduced $\chi^2 \sim 1.6$, 
the model described the data well at a few percentage level and 
most of the discrepancies are at low energies. The high energy photon index is not too sensitive to
the actual model used as we show later when we fit only the high energy part ($3$-$10$ keV) of the spectra 
and obtain qualitatively similar results. Nevertheless, we caution against over-interpretation
of the best fit values obtained in these fits, especially for the components affecting the low
energy part of the spectra.

{
\section{The effect of complex absorption and reflection component}}

{
The spectra of Mrk~335, may be affected by complex absorption and there may be a relativistically
blurred reflection component which may dominate at low and high energies. Given the quality of the data many of these
complex models may be degenerate, however in the present context it is useful to quantify the effect of these models
on the high energy spectral index. The motivation here is not to obtain a physically self consistent model but rather
to understand their effect of the high energy spectral index and hence in the primary result of this work.}

{
To take into account the possibility of complex absorption, we include in the phenomenological model, partial ionised
absorber represented by the XSPEC routine ``zxipcf'' \citep{Mil07}. The absorption is characterised by three parameters namely
the column density, the covering fraction and the ionisation parameter of the absorber. For the addition
of three parameters the changes in $\chi^2$ for the four high flux observations with IDs 0101040101,0306870101,  0600540601
 and 0600540501  were 12.7, 22.0, 3.1, 9.0 and 12.0 respectively. More importantly, the best fit high energy
spectral indexes obtained were $2.05 \pm 0.05$, $1.78\pm 0.06$, $1.75\pm 0.05$  and $1.81\pm 0.02$. When compared
with the high energy spectral indexes obtained from the phenomenological model, one can see that  the errors
are larger and any change in the index is less than $0.1$. In contrast, for the deep low flux state, ID 0510010701,
the $\Delta \chi^2 = 45$ and the spectral index obtained was $1.55\pm 0.15$ which is significantly different from
the phenomenological model.}

{
Next we consider the possible effect of a complex relativistically blurred reflection component in the spectra. Since,
the ionised reflection component produces a complex soft excess which needs to be modelled with absorption, we limit
our analysis to energies $> 2$ keV. This is adequate since our interest here is to study the effect of the component
on the high energy index. Since for the phenomenological model, the spectra shows the presence of a narrow and broad
Iron lines, we consider a power-law and two reflection components represented by the table model ``reflionx'' \citep{Ros05}.
For one of the 
the reflection component we convolve it  using the
relativistic blurring  model ``kdblur'' where we fix the inner and outer radii, but allow for the emissivity
index to be free. The best fit power-law index for the four high flux state observations with 
 IDs 0101040101,0306870101,  0600540601 and 0600540501 were $2.09 \pm 0.05$, $2.16\pm 0.02$, $1.77\pm 0.03$  and $1.64\pm 0.05$.
We note that for the observation with ID. 0306870101 the change in spectral index as compared to the
phenomenological model is $\sim 0.28$ while for the others it is $< 0.15$. For the deep low flux state, ID 0510010701,
the spectral index obtained was $3.00\pm .07$ which is radically different than the value obtained earlier.}

{
Thus, the effect of complex absorption or reflection is dramatic for the deep low state. This is expected
based on the analysis of \citet{Grupe et al.08a}. For the other higher flux states, complex absorption may change the index
by $< 0.1$ from those of the phenomenological model. The effect of Complex reflection on the index is also
modest $< 0.15$ but we note that it may be large $\sim 0.28$ for the observation with  ID. 0306870101, which
incidentally is also the highest flux state in the sample.}

\section{Flux-resolved spectroscopy}

\begin{figure}
\includegraphics[width=6cm, angle=270]{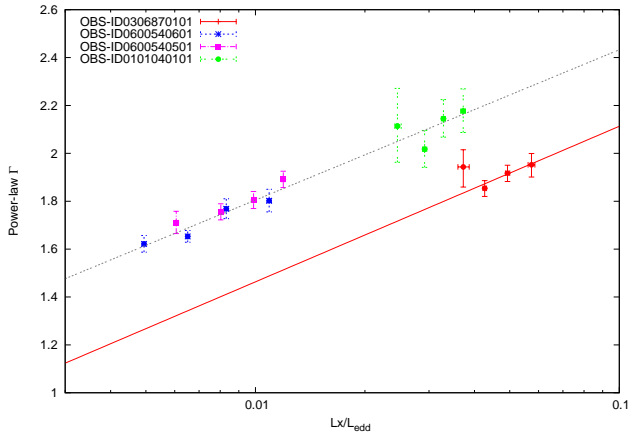}
\caption{ The high energy photon index, $\Gamma$ versus the {X-ray} Eddington ratio for Mrk~335.
The {X-ray} Eddington ratio is $L{_{X}}/L_{Edd}$ where $L{_{X}}$ is the unabsorbed luminosity in the $0.3$-$10$ keV range. 
The solid line  is a fit to three of the data sets and has a slope  $m = 0.64\pm 0.04$ and
intercept $c = 3.08\pm 0.08$ with a reduced $\chi^{2}= 0.57$. The bottom dotted straight line is a fit only to the data set (ID0306870101) and has a slope  $m = 0.65\pm 0.04$ and
intercept $c = 2.76\pm 0.07$ with a reduced $\chi^{2}= 0.76$.}
\label{mrk335}
\end{figure}

\begin{figure}
\includegraphics[width=6cm, angle=270]{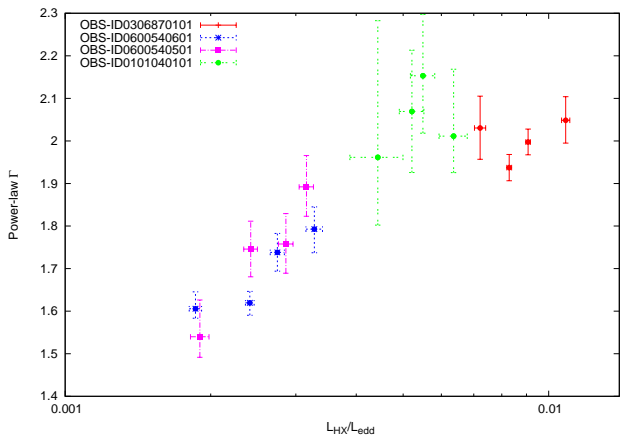}
\caption{ Same as Figure \ref{mrk335} except that only the energy range $3$-$10$ keV is taken
into account. The spectra are fitted with a power-law and Iron line. The luminosity $L{_{HX}}$ corresponds to the energy range $3$-$10$ keV. Note the similarity with Figure \ref{mrk335} which implies that the qualitative results are not sensitive to the spectral model adopted.  }
\label{mrk335HE}
\end{figure}

The light curves for each of the observations {of Mrk 335 and Ark 564 are shown in Figure \ref{lc1} and Figure \ref{lc2}} which reveals
that for nearly all observations there is significant flux variation in timescales of
$\sim 10^4$ s. To investigate the variation of the photon index, each observation was
split into 3 or 4 flux levels (marked by horizontal dotted lines in Figures \ref{lc1} and \ref{lc2}) and the corresponding spectrum was generated.  {We do not consider the 2007 observation of Mrk~335 (ID0510010701) for the flux resolved
spectroscopy since as shown in the previous section the spectra is clearly  complex and the phenomenological model
used here is not adequate.}

Each of the flux level spectra was fitted with the same phenomenological model used for the
average spectra. However, due to the lower statistics of the flux resolved spectra,
there were several parameters which were either not constrained or were consistent
within error bars to be not varying during the observations. Thus, several
spectral parameters were fixed to their best fit values obtained from the
fitting of the average spectra. These were, the energies of the different edges and
emission lines, the temperature and normalisation of the soft photon Comptonization (i.e. $kT_{bb}$ and 
normalisation of the ``nthcomp'' component) and the normalisation and emissivity index of the Iron line
(i.e. $\beta$ and normalisation of the ``diskline'' component).   

For each flux resolved spectra, the unabsorbed flux in the $0.3$-$10$ keV band
was computed using the XSPEC model ``cflux''. The fluxes were converted into
luminosities, $L$ by using the luminosity distances of 103 Mpc for Mrk~335 and
98.5 Mpc for Ark~564, which are the average values quoted in the NED website. 
 For Mrk~335, the black hole mass
was taken to be 1.4 $\times$10$^{7}$~M$_{\odot}$ \citep{Peterson et al.2004,
Grier et al.2012} while for Ark~564 we adopt a value of
2.61 $\times$ 10$^{6}$~M$_{\odot}$ \citep{Botte et al.2004} to obtain their
respective Eddington luminosities $L_{Edd}$.{ The best fit parameters for the
flux resolved spectra are listed in Table \ref{spect}}.

  {In this work we consider the {X-ray} Eddington ratio i.e. $L{_{X}}/L_{Edd}$ instead
of the Bolometric one and discuss the implication of this in the last section.}
In Figure \ref{mrk335}, the high energy photon index is plotted against the
{X-ray} Eddington ratio for the different flux resolved spectra. 
Three of the observations cover an order of magnitude range in {X-ray} Eddington
ratio from $0.004$ to $0.04$ and show a tight correlation between the
photon index and {X-ray} Eddington ratio.  A straight line fit to these three observation sets gives
a slope  $m = 0.64\pm 0.04$, intercept   $c = 3.08\pm 0.08$ and reduced $\chi^{2}= 0.57$.
The observation with the highest {X-ray} Eddington ratio $> 0.04$, does not follow the linear
relation. Instead fitting the flux resolved spectra for that observation only, one
gets a straight line with  slope $m = 0.65\pm 0.04$, intercept $c = 2.76\pm 0.07$ 
and reduced $\chi^{2}= 0.76$. In other words, the data follows a parallel track.
Note that the lowest luminosity observation ($L{_{X}}/L_{Edd} \sim 0.002$) was not included
in the straight line fit. Nevertheless, the fit to the high flux data, passes through
the lowest flux one.

To verify  that the results do not depend sensitively on the phenomenological spectral
model used, we repeat the flux resolved spectroscopy analysis using only the
energy band $3$-$10$ keV. We fit this energy band with a power-law and 
a broad Iron line. Figure \ref{mrk335HE} shows the results of this analysis
where the photon index is plotted against $L{_{HX}}/L_{Edd}$, where $L{_{HX}}$ is
computed using the unabsorbed flux in the $3$-$10$ keV band.

\begin{figure}
\includegraphics[width=6cm,angle=270]{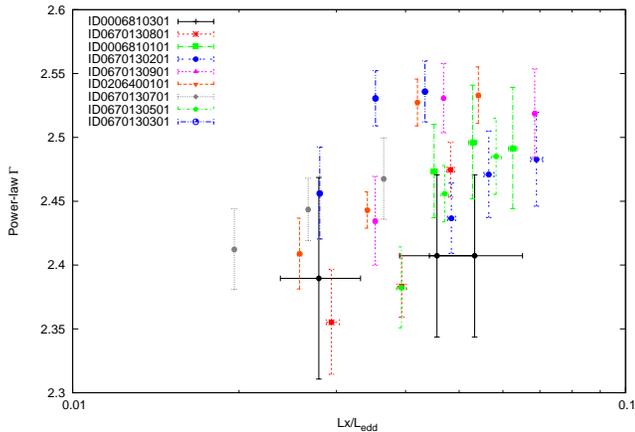}
\caption{The high energy photon index, $\Gamma$ versus the {X-ray} Eddington ratio for Ark~564.
The {X-ray} Eddington ratio is $L{_{X}}/L_{Edd}$ where $L{_{X}}$ is the unabsorbed luminosity in the $0.3$-$10$ keV range. }
\label{ark564}
\end{figure}

Figure \ref{ark564} shows the results of the flux resolved analysis 
(for the complete $0.3$-$10$ keV band) for Ark~564. Here, despite the
larger number of observations, the flux variation is modest
with the {X-ray} Eddington ratio ranging from $0.02$-$0.07$. Although there is
more scatter than the case of Mrk~335, there is a correlation between the
spectral index and {X-ray} Eddington ratio. A straight line fit gives a
slope $m = 0.22\pm 0.08$, intercept $c = 2.75\pm 0.1$ 
and a large reduced $\chi^{2}= 2.75$. From the Figure, it seems that
for the flux resolved spectra of each individual observation the
correlation may be better and perhaps the different observations are
parallelly shifted with respect to each other. However, the
statistics is not good enough to make any concrete statements.

\begin{figure}
\includegraphics[width=6cm, angle=270]{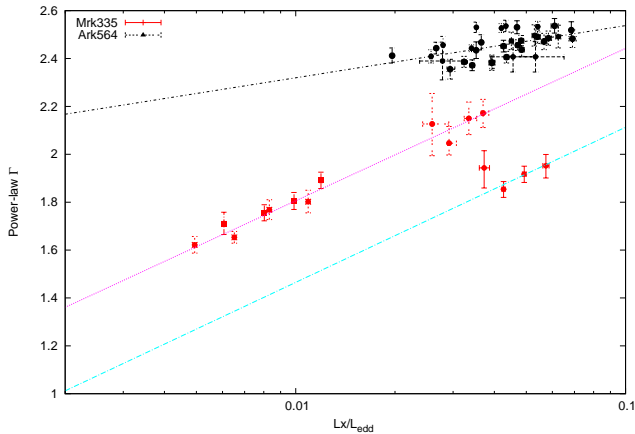}
\caption{ Summary figure for the high energy photon index, $\Gamma$ versus the {X-ray} Eddington ratio for both Mrk~335 and Ark~564. The {X-ray} Eddington ratio is $L{_{X}}/L_{Edd}$ where $L{_{X}}$ is the unabsorbed luminosity in the $0.3$-$10$ keV range. For Mrk~335 two parallel tracks are observed with a possible switching at $L{_{X}}/L_{Edd} \sim 0.04$, while for Ark~564 the correlation is flatter and with significantly
more scatter. } 
\label{Tot}
\end{figure} 

\section{Discussion}

Figure \ref{Tot} summarises the results of the work by showing the variation of the
high energy photon index versus the {X-ray} Eddington ratio for both Mrk~335 and Ark~564.
For Mrk~335, there are two parallel tracks with the possibility that the sources shifts
to the lower track for $L{_{X}}/L_{Edd}> 0.04$. For Ark~564 the points are clustered at a larger
$\Gamma$ and while there is a positive correlation, the slope is flatter and points show
much more scatter than the points for Mrk~335.

It should be noted that the luminosity used in this work is in the energy
range $0.3$-$10$ keV and is not the bolometric one and hence these results
cannot be compared directly with those where the bolometric luminosity or
X-ray luminosity in some other energy range has been used. {The conversion to Bolometric 
luminosity often involves several uncertain factors and hence is avoided in this work.}
Since for Mrk~335,
we find the the index correlates well with the X-ray luminosity for nearly an order of
magnitude change in luminosity, this may indicate that over this range the
bolometric correction is nearly constant. On the other hand, it may well be
that for this source the X-ray index correlates better with the X-ray luminosity and not with 
the bolometric one. The uncertainties and model dependency of estimating the
bolometric luminosity (including the general non-availability of simultaneous
multi-wavelength data) does not allow for concrete statements. For Mrk~335, there
is only one observation for which the X-ray Eddington ratio is larger than 0.04 and the
source follows a lower parallel track. It is necessary to confirm this behaviour with
future observations when the source is equally luminous. It might be that there is
a real  transition at the X-ray Eddington ratio $\sim 0.04$ or that the source can actually follow
any of the two tracks at the same  luminosity. 

In the Comptonization context, the high energy power-law index is inversely 
proportional to the Compton Amplification factor $A$, which is the ratio
of the luminosity of Comptonizing cloud, $L_c$ to the input soft photon
luminosity $L_{inp}$ i.e. $A = L_c/L_{inp}$. The input luminosity $L_{inp}$
depends on the luminosity of the soft photon source and the fraction of photons
which enter the Comptonizing region. The latter depends on the accretion geometry
of the system. Thus, the correlation between the photon index and the observed luminosity
could occur if $A$ decreases with luminosity or in other words $L_c$ varies less
rapidly with the observed luminosity than $L_{inp}$. The shift to the lower  parallel track can
be explained if there is a decrease in $L_{inp}$ at the X-ray Eddington ratio of $\sim 0.04$, perhaps
caused by a change in the fraction of photons entering the Comptonizing region. This would
mean that the accretion geometry for the two parallel tracks are different.

The high energy photon index could also be affected by the presence of a strong
reflection component, especially if the reflection is from partially ionised matter
and/or is relativistically blurred such that it has a significant contribution 
to the spectra below $10$ keV. The reflection component would tend to flatten the
high energy spectrum and hence the correlation may be caused  by the reflection
component decreasing as the source becomes more luminous. 
Indeed, the anomalous 
low luminosity observation of Mrk~335 of  July 2007, can be modelled as
being dominated by a blurred reflection component \citep{Grupe et al.08a}.
It may also be that the primary correlation between the index and luminosity
is due to variation of the Compton Amplification factor, while the shift to
a different parallel track is due to appearance of a strong reflection component.
The appearance of a stronger reflection component would also mean that there
was a change in the accretion disk geometry. Thus, a qualitative change in the
accretion disk geometry seems to be required to explain the parallel tracks
seen for Mrk~335. Changes in the accretion disk
geometry at the same X-ray luminosity seems to indicate that the accretion flow
is not determined only by the local accretion rate but rather it may also
depend on the previous history of the accretion rate variation. 
{Our brief analysis of including a blurred reflection component suggests
that this indeed may be the case. It is particularly interesting to note that for the
highest flux observation (ID 101040101) a reflection component may increase the
index by $0.28$ while a more modest variation is expected for the others (Section 4)}. 
 { If indeed this highest flux observation has a stronger reflection component and its spectral
index is more close to $\sim 2$ then  it would seem that the index has a positive correlation with flux but
saturates to $\sim 2$ at high Eddington ratios. Interestingly a similar behavior is observed in
Fig. 5, where the results for a simple power law plus Iron line fit to
the 3-10 keV energy range are shown.
However, this can be confirmed only with broad band data covering energies $> 10$ keV  {where the
reflection component will be better constrained.}

The results for Ark~564 are significantly different than that for Mrk~335, with
a flatter correlation and more scatter. This suggests that the behaviour of AGNs is
not homogeneous and perhaps cannot be understood by studying a large sample of them.
Long term sensitive and broad band  monitoring of individual AGNs may be expected to
provide better insights.

\section*{Acknowledgments}

RS acknowledges IUCAA visitor program and AP acknowledges seed
money grant from Tezpur University and visiting  associateship
IUCAA, Pune. This research made use of data obtained from the High
Energy Astrophysics Science Archive Research Center (HEASARC),
provided by NASA's Goddard Space Flight Center. The authors would
like to thank the anonymous referee for useful suggestions which has
enhanced the quality of the paper.

\begin{table*}

\begin{center}
\caption{Spectral parameters for each flux state (for Mrk~335 \& Ark~564)
derived from different \textit{XMM-Newton} observations in the (0.3-10.0)keV range. The full version is available to download online}
\label{spect}
\begin{tabular}{lcccccccccc}
\hline
\hline
Obs ID &flux state& \multicolumn{1}{c}{zedge1}&\multicolumn{1}{c}{zedge2}&\multicolumn{1}{c}{zedge3}&\multicolumn{2}{c}{Simpl}&\multicolumn{1}{c}{nthComp}&red $\chi^{2}$/d.o.f.\\ &&$\tau$&$\tau$&$\tau$&$\Gamma$&FracScat&$\Gamma$\\
0101040101&1&  0.002$\times$10$^{-4}$$_{-  0.002 }^{+  908.7}$ &$  0.073_{-  0.065 }^{+  0.101}$ &-&$  2.114_{-  0.151 }^{+  0.157}$ &$  0.140_{-  0.030 }^{+  0.039}$ &$  2.979_{-  0.190 }^{+  0.125}$ &0.92/ 82\\

(Mrk~335)&2&$  0.132_{-  0.055 }^{+  0.055}$ &$  0.024_{-  0.024 }^{+  0.055}$ &- &$  2.017_{-  0.076 }^{+  0.078}$ &$  0.120_{-  0.013 }^{+  0.016}$ &$  2.780_{-  0.084 }^{+  0.096}$ &1.01/120\\
&3&$  0.098_{-  0.055 }^{+  0.055}$ &$  0.074_{-  0.053 }^{+  0.053}$ &- &$  2.145_{-  0.077 }^{+  0.080}$ &$  0.140_{-  0.017 }^{+  0.020}$ &$  2.921_{-  0.106 }^{+  0.127}$ &0.89/120\\
&4&$  0.077_{-  0.063 }^{+  0.063}$ &$  0.070_{-  0.061 }^{+  0.061}$ &- &$  2.177_{-  0.090 }^{+  0.093}$ &$  0.160_{-  0.022 }^{+  0.028}$ &$  2.896_{-  0.127 }^{+  0.162}$ &1.05/112\\
\hline

  0006810101&1&$  0.107_{-  0.041 }^{+  0.041}$ &$  0.121_{-  0.046 }^{+  0.046}$ &$ -0.006_{-  0.036 }^{+  0.037}$ &$  2.473_{-  0.036 }^{+  0.037}$ &$  0.189_{-  0.012 }^{+  0.013}$ &$  2.401_{-  0.076 }^{+  0.085}$ &0.93/137\\
(Akn~564)&2&$  0.093_{-  0.048 }^{+  0.048}$ &$  0.127_{-  0.054 }^{+  0.054}$ &$  0.041_{-  0.044 }^{+  0.044}$ &$  2.496_{-  0.044 }^{+  0.045}$ &$  0.197_{-  0.015 }^{+  0.017}$ &$  2.289_{-  0.083 }^{+  0.093}$ &0.99/130\\
&3&$  0.124_{-  0.052 }^{+  0.051}$ &$  0.199_{-  0.059 }^{+  0.059}$ &$  0.045_{-  0.047 }^{+  0.047}$ &$  2.491_{-  0.047 }^{+  0.048}$ &$  0.204_{-  0.016 }^{+  0.018}$ &$  2.177_{-  0.082 }^{+  0.092}$ &1.05/126\\                                 
\hline

\end{tabular} 
\end{center}
\end{table*}

\label{lastpage}
\end{document}